\documentclass[prl,twocolumn]{revtex4}
\usepackage{mathrsfs}
\usepackage{amssymb}
\usepackage{amsmath}
\usepackage{amsfonts}
\usepackage{graphicx}           
\usepackage{rotating}    
\usepackage{dcolumn}
\usepackage{epsfig}

\begin{document}

\title{Detecting phase synchronization by localized maps: \\
  Application to neural networks} 
\author{T.  Pereira, M. S. Baptista, and J.  Kurths}
\affiliation{ Universit{\"a}t Potsdam, Institut f{\"u}r Physik Am Neuen
  Palais 10, D-14469 Potsdam, Deutschland} 

\date{\today}

\begin{abstract}
  
  We present an approach which enables to state about the existence of
  phase synchronization in coupled chaotic oscillators without having
  to measure the phase.  This is done by observing the oscillators at
  special times, and analyzing whether this set of points is
  localized.  In particular, we show that this approach is fruitful to
  analyze the onset of phase synchronization in chaotic attractors
  whose phases are not well defined, as well as, in networks of
  non-identical spiking/bursting neurons connected by chemical
  synapses.
  
\vglue 0.4 truecm

%PACS: 05.45.Xt, 05.45.-a, 05.45.Pq, 02.50.CW

\end{abstract}

\maketitle

Many neural networks rely on a synchronous behavior for a proper
functioning, e.g. information transmission \cite{bondarenko,borisyuk},
pattern recognition \cite{haken}, and learning \cite{chao}.
Nevertheless, the onset of synchronization in a network may also lead
to some diseases as  Parkinson disease \cite{parkinson} and
epilepsy \cite{mormann:2003}.  Studies on this topic have been
concentrated on synchronization of networks of identical chaotic
oscillators, in which the onset of complete synchronization takes
place \cite{pecora}.  However, they do not treat the onset of
synchronization in networks of non-identical chaotic oscillators, case
typically found in nature \cite{reviews}, where complete
synchronization is much harder to be achieved. Indeed, in such
networks a weaker kind of synchronization may take place, the phase
synchronization (PS).  The condition for PS between two subsystems $k$
and $j$ can be written as:
\noindent
\begin{equation}
|\phi _{k}(t)-r\phi _{j}(t)|<c,  \label{PS}
\end{equation}
\noindent where $\phi _{k,j}$ is the phase of the subsystem
$\Sigma_{k,j}$, $c\in \mathbb{R} $ is a constant, and $r$ is a
rational number.  PS is a common phenomenon in interacting chaotic
oscillators \cite{reviews}, and plays a major role in physical
processes linked to communication \cite{Murilo-Canal}, and
communication processes in the human brain
\cite{fell:2002,mormann:2003}.

In order to state about the existence of PS, one has to introduce a
phase $\phi(t)$ for the chaotic oscillator, what is not
straightforward.  Indeed, there is no general definition of phase to
chaotic attractors, and depending on the situation one has to decide
which phase is more suitable. In fact, in some oscillators it is
rather unclear which phase one should use, especially in non-coherent
oscillators with more than one time scale, typically found in neuronal
dynamics with bursting/spiking behavior.

In this letter, we present a general and easy way to identify PS
without having to access explicitly the phase.  The approach consists
in defining maps, which are a natural extension of the stroboscopic
map, to coupled chaotic oscillators, in which the oscillators are
observed at especial times.  PS implies the existence of maps of the
attractor that appear as localized structures in the accessible
phase-space.  The fact that PS produces subsets of the attractor that
are localized structures, by particular observations was previously
used as a way to detect PS in chaotic oscillators
\cite{reviews,Non-Transitive,epa}. Here, we extent these results by
demonstrating that localized sets can be constructed while in PS by
means of {\it any typical physical} observation, which has a strong
impact in the field of experimental physics, since in the laboratory
measurements are restricted to the limitations of the experiment. Note
that since this approach does not requite any further calculation, but
just the analysis of whether the sets are localized, it can be used in
real time experiments for PS detection.  We illustrate the power of
this approach by analyzing PS in a network of non-identical
Hindmarsh-Rose (HR) spiking/bursting neurons connected via chemical
synapses.

The classic stroboscopic map is defined in periodically driven chaotic
oscillators.  It consists in sampling the chaotic trajectory at times
$n T_0$, where $n$ is an integer and $T_0$ is the period of the
driver.  The stroboscopic map was used to detect PS
\cite{reviews,epa}.  The basic idea is that if the stroboscopic map is
localized in the attractor, PS is present. To generalize the
stroboscopic map to coupled chaotic oscillators, we do the following:
Given two subsystems $\Sigma_k$ and $\Sigma_j$, we observe $\Sigma_k$
at times when some event in the oscillator $\Sigma_j$ happens. As a
consequence of these observations, we get a discrete set
$\mathcal{D}_k$.  Then, we demonstrate that if there is PS then the
set $\mathcal{D}_k$ is localized.

\begin{figure}[!h]
  \centerline{\hbox{\psfig{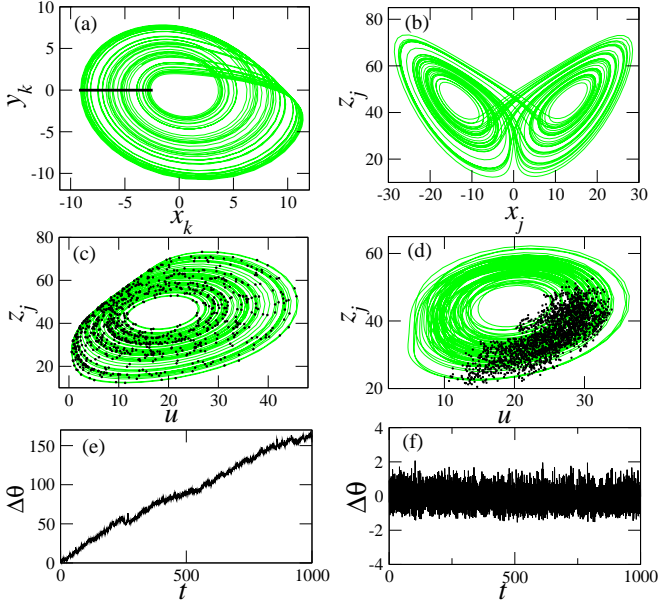}}}
\caption{\label{NovaFig} PS onset between a Lorenz oscillator driven by a R\"ossler
  oscillator. The attractor of the oscillators are depicted in gray
  (a-d). In (a) we show the projection of the R\"ossler attractor into
  the subspace $(x_k,y_k)$. The black line shows the Poincar\'e
  section at $y_k=0$ with the constrain $\dot{y}_k<0$.  The projection
  of the Lorenz attractor in the subspace $(x_j,z_j)$ is depicted in
  (b), and in (c,d) into the subspace $(u,z_j)$.  For $\epsilon=0.0$,
  the set $\mathcal{D}_j$ (in black) spreads over the attractor of the
  oscillator $\Sigma_j$ (c), and there is no PS; the phase difference
  diverges (e). For $\epsilon=13.0$ the oscillators present PS. The
  set $\mathcal{D}_j$ is localized, it does not fulfill the attractor
  of $\Sigma_j$ (d); the phase difference is bounded (f).}
\end{figure}

In order to introduce our ideas, we analyze PS in two coupled chaotic
oscillators, namely the Lorenz oscillator driven by the R\"ossler.
Further, we extent this result to general compact oscillators. The
subsystem $\Sigma_k$ corresponds to the R\"ossler oscillator and
$\Sigma_j$ to the Lorenz. They are coupled unidirectionally in the
driver response scheme.  An event in the $\Sigma_k$ is considered to
happen when its trajectory crosses a Poincar\'e section $y_k=0$.  As a
result, we get the series of times {\small
  $(\tau^i_k)_{i\in\mathbb{N}}$}, where $\tau^i_k$ is the time at
which the $i$th crossing of the trajectory of $\Sigma_k$ occurs in a
Poincar\'e plane.  The two coupled oscillators are given by:
{\small
$$
\begin{array}{lllll}
\dot{x}_{k} = -\alpha(y_{k}+z_{k}) & ~&~ &  \,\,\dot{x}_{j} = \sigma(y_{j}-x_{j}) + \epsilon(x_k - x_j) &  ~ \\
\dot{y}_{k} = \alpha(x_{k}+0.2y_{k}) & ~& ~&  \,\,  \dot{y}_{j} = rx_j - y_j -x_jz_j & \,\,\,~ \\
\dot{z}_{k} = \alpha[0.2+z_{k}(x_{k}-5.7)] & ~ &~ &  \,\, \dot{z}_{j} = x_jy_j - \beta z_j & ~ \\ 
\end{array}
$$}
\noindent
with $\alpha=13$,$\sigma=16$, $r=45.92$, and $\beta=4$. Since the
trajectory of the R\"ossler oscillator rotates around a fixed point
[Fig.  \ref{NovaFig}(a)], we can define a phase $\theta_k =
tan^{-1}(y_k/x_k)$ which provides: $\theta_k(t) = \int_0^t (\dot{y}_k
x_k - \dot{x}_k y_k)/(x^2_k + y^2_k) dt$. The trajectory of the Lorenz
does not have an unique center of rotation, see Fig. \ref{NovaFig}(b).
However, if we consider the projection $(u,z_j)$ with {\small
  $u=\sqrt{x_j^2+y_j}$}, the trajectory projected into this subspace
presents an unique center of rotation.  Thus we also define a phase
$\theta_j = tan^{-1}[(z_j-z_{j0})/(u-u_0)]$, where $(u_0,z_{j0})=(19,45)$ is
the center of rotation in the subspace $(u,z_j)$, which provides  
$\theta_j(t) = \int_0^t [\dot{z}_j u - \dot{u} (z_k - z_{j0})]/[(u-u_0)^2 + (z_j-z_{j0})^2] dt$

For $\epsilon = 0.0$, we construct the set $\mathcal{D}_j$ by sampling
the trajectory of $\Sigma_j$ at times $\tau_k^i$. The set
$\mathcal{D}_j$ spreads over the trajectory of $\Sigma_j$; there is no
PS, the phase difference $\Delta \theta = \theta_k - \theta_j$
diverges [fig.  \ref{NovaFig} (c,e)]. Indeed, a calculation of the
frequencies shows that $\langle \dot{\theta}_k \rangle \approx 13.94$
and $\langle \dot{\theta}_j \rangle \approx 13.75$.  As we increase
the coupling, PS appears. In particular, for $\epsilon = 13.0$ the set
$\mathcal{D}_j$ is localized, and the phase difference is bounded
[fig.  \ref{NovaFig}(d,f)]. The average frequencies are $\langle
\dot{\theta}_k \rangle = \langle \dot{\theta}_j \rangle \approx
13.95$.
\noindent
\noindent

Next, we demonstrate that the sets $\mathcal{D}_j$ of the attractor
that appear as localized structures imply PS, and vice-versa. We first
show for a Poincar\'e section for a better understanding of the ideas,
and then we generalize these results to any possible event.
$\Sigma_j$ is given by the dynamical system $\dot{{\bf x}}_j = {\bf
  G}_j({\bf x}_j)$, let $F_j^t$ be the flow, $\Gamma_j$ the Poincar\'e
section, and $\Pi_j$ the Poincar\'e map associated to the section
$\Gamma_j$, such that given a point ${\bf x}_j^{i} \in \Gamma_j$, so
${\bf x}_j^{i+1} = \Pi_j({\bf x}_j^{i})$ = $F_j^{\Delta
  \tau_j^{i+1}}({\bf x}_j^i)$, where $\Delta \tau_j^{i}$ = $\tau_j^{i}
- \tau_j^{i-1}$.  From now on, we use a rescaled time $t^{\prime} = t
/ \langle T_j \rangle $.  For a slight abuse of notation we omit the
$\prime$.  The average return time is given by $\langle T_k \rangle =
\sum_{i=0}^{N} \Delta \tau^i_k / N = \tau_k^N / N$, and the time is
rescaled, such that $\langle T_k \rangle =1$. From our hypothesis that
both oscillators present phase-coherent oscillations, there is a
number $\kappa_k$ such that $| \tau _k ^{i} - i \langle T_k \rangle |
\leq \kappa_k,$ where $\kappa_k \ll 1$. If both oscillators are in PS,
then $\langle T_k \rangle = \langle T_j \rangle$, and so:
\begin{equation}
|  \tau_k^{i} -  \tau_j^{i}| \leq \tilde \kappa,
\label{diferenca_temporal}
\end{equation}
\noindent 
with $\tilde \kappa \le \kappa_k + \kappa_j \ll 1 $.  Now, we analyze
one typical oscillation, using the basic concept of recurrence. Given
the following starting points ${\bf x}_k^0 \in \Gamma_k$ and ${\bf
  x}_j^0 \in \Gamma_j$, we evolve both until ${\bf x}_j^0$ returns to
$\Gamma_j$.  Let us introduce $\Delta \tau^i = \Delta \tau_k^i -
\Delta \tau_j^i$.  So, ${\bf F}_j^{\Delta \tau_j^{1}}({\bf x}_j^0) =
\Pi_j({\bf x}_j^0)= {\bf x}_j^1 \in \Gamma_j $.  Analogously, ${\bf
  F}_k^{\Delta \tau _j^{1}}({\bf x}_k^0) = {\bf F}_k^{ \Delta
  \tau_k^{1} + \Delta \tau^1}({\bf x}_k^0) = {\bf F}_k^{ \Delta
  \tau^1} \circ {\bf F}_k^{ \Delta \tau_k^{1}}({\bf x}_k^0) = {\bf
  F}_k^{ \Delta \tau^1}( \Pi_k({\bf x}_k^0)) = {\bf F}_k^{ \Delta
  \tau^1}({\bf x}_k^1)$.  Now, by using the fact that $| \Delta \tau^i
| < \tilde \kappa $, we can write: ${\bf F}_k^{ \Delta \tau^1}({\bf
  x}_k^1) \approx {\bf x}_k^1 + {\bf G}({\bf x}_k^1) \tilde \kappa +
O(\tilde \kappa ^2).$ So, given a point ${\bf x}_k \in \Gamma_k$
evaluated by the time when the trajectory of $\Sigma_j$ returns to the
section $\Gamma_j$, the point ${\bf x}_k$ returns near the section
$\Gamma_k$, and vice-versa.  For a general case, we have to show that
a point, in the section $\Gamma_k$, evolved by the flow for an
arbitrary number of events in the oscillator $\Sigma_j$, still remains
close to $\Gamma_k$.  But, this is straightforward, since
$|\sum_{i=0}^N \Delta \tau^i| = |\tau_k^N - \tau_j^N | < \tilde
\kappa$. So, we demonstrated that the PS regime implies the
localization of the set $\mathcal{D}_k$.  Now, we show that the
localization of the set $\mathcal{D}_k$ implies PS.  Supposing that we
have a localized set $\mathcal{D}_k$, so, eq.
(\ref{diferenca_temporal}) is valid, by the above arguments.
Therefore, we just have to show that eq.  (\ref{diferenca_temporal})
implies PS.  To do so, we note that at every crossing of the
trajectory with the Poincar\'e section the phase increases $2\pi$, as
a consequence $\phi_k(\tau_k^i)= 2 i \times \pi$. Then,
$|\phi_k(\tau_k^i) - \phi_j(\tau_k^i)| = |\phi_k(\tau_k^i) -
\phi_j(\tau_j^i + \zeta )|$, where $\zeta = \tau_k^i - \tau_j^i$. Now,
expanding the phase $\phi_j$ in Taylor series around $\tau_j^i$, we
have $\phi_j(\tau_j^i + \zeta ) \approx 2 i \pi +
\dot{\phi}_j(\tau_j^i) \times \zeta + \mathcal{O}(\zeta^2)$, as a
result, the phase difference can be written as :
\noindent
\begin{equation}
|\phi_k(\tau_k^i) - \phi_j(\tau_k^i)| \le \Lambda \times |\tau_k^i - \tau_j^i | \le \Lambda \times \tilde \kappa, 
\label{timeImphase}
\end{equation}
\noindent
where, $\Lambda = max_{t,j}\{\dot{\phi}_j(t)\}$. Therefore, we showed
that boundness in eq.  (\ref{diferenca_temporal}) implies a bound in
the phase difference at times $\tau_k^i$.  However, since the phase
depends smoothly on time, and the Poincar\'e section can be smoothly
changed, the boundness in eq. (\ref{timeImphase}) also holds at the
continuous time.  Thus, we conclude our result.
 
An important point to stress is that it is not always possible to
define a Poincar\'e Section on the attractor in such a way that a
phase increases $2\pi$ every crossing. As an example we quote the
non-coherent attractors with no proper rotations, where the definition
of such section is not possible. Moreover, even if the oscillators are
coherent, it might happen that the accessible data is not suitable to
define a section, but rather to measure the entrance of the trajectory
in some small region of the phase space. That does not constitute a
problem, because PS implies localization of the set $\mathcal{D}_k$,
independently on the event definition.  

Let us first discuss the idea of localization.  If the set
$\mathcal{D}$ is a subset of $\Phi$, we say that $\mathcal{D}$ is
localized (with respect to $\Phi$) if there is a cross section $\Psi$
and a neighborhood $\Lambda$ of $\Psi$, such that $\mathcal{D} \cup
\Lambda = \emptyset$.  In particular, for practical detections, one
may check whether $\mathcal{D}$ is localized, by the following
technique. If there is PS, for ${\bf y} \in \mathcal{D}$ it exists
infinitely many ${\bf x} \in \Lambda$ such that ${\bf y} \cap
B_{\ell}({\bf x}) = \emptyset,$ where $B_{\ell}({\bf x})$ is an open
ball of radius $\ell$ centered at the point ${\bf x}$, and $\ell$ is
small. Then, we may vary ${\bf y},{\bf x}$ (one may take ${\bf x}$ to
be an arbitrary point of the attractor) and $\ell$, to determine
whether $\mathcal{D}$ is localized.

The event definition that generates the time series $\{ \tau_k^i\}_{i
  \in \mathbb{N}}$ can be arbitrary. Therefore, the event could be a
local maximum/minimum, the crossing of a dynamical variable with a
threshold, the entrance in an $\varepsilon$-ball, and so on.  The only
constraint is the event must be typical.  We also suppose that
there is a function phase $\phi_k$, in such a way that $\dot{\phi}_k =
\Omega_k$, where $\Omega_k$ is continuous and $\Omega_k \le \Upsilon$.
Under such hypotheses, we can state that: {\it Given any typical event,
  with positive measure, in the oscillator $\Sigma_{k,j}$, generating
  the times $(t_{k,j}^i)_{i\in \mathbb{N}}$, if there is PS the
  observation of $\Sigma_k$ at $(t_j^i)_{i\in \mathbb{N}}$ generates a
  localized set $\mathcal{D}_k$.}

Next, we demonstrate this result, doing so we extent these ideas to
non-coherent oscillators.  The strategy to demonstrate the previous
results to an arbitrary event is the following:$(i)$ Note that the
phase $\phi_j(t)$ naturally defines a section, namely $\tilde
\Gamma_j$, in the attractor such that at the $N$th crossing of the
trajectory of $\Sigma_j$ with $\tilde \Gamma_j$ the phase is equal to
$N \times 2 \pi$. Obviously, this section depends on the initial
conditions. $(ii)$ Suppose that we construct the set $\mathcal{D}_k$
by observing the trajectory of $\Sigma_k$ at every crossing of the
trajectory of $\Sigma_j$ with $\tilde \Gamma_j$. Then, following the
previous results, PS implies the localization of $\mathcal{D}_k$, and
vice-versa.  $(iii)$ Suppose that we have a small piece $P_{\tilde
  \Gamma_j}$ of the section $\tilde \Gamma_j$, such that the crossings
of the trajectory of $\Sigma_j$ with $P_{\tilde \Gamma_j}$ produces a
subsequence $(\tau_j^{n_i})_{n_i\in\mathbb{N}}$ of the sequence $(
\tau_j^i)_{i\in\mathbb{N}}$. Thus, we just note that if the
observation of the trajectory of $\Sigma_k$ at times
$(\tau_j^i)_{i\in\mathbb{N}}$ gives place to a localized set
$\mathcal{D}_k$, the observation at times
$(\tau_j^{n_i})_{n_i\in\mathbb{N}}$ also gives place to a localized
set $\tilde{\mathcal{D}}_k$ which is a subset of $\mathcal{D}_k$.
Therefore, we showed that the observation of the trajectory of
$\Sigma_k$, when the trajectory of $\Sigma_j$ returns to $P_{\tilde
  \Gamma_j}$, also leads to a localized set in $\Sigma_k$.  $(iv)$
Next, we show that an event does not have to be a piece of the section
$\tilde \Gamma_j$ in order to obtain a localized set in $\Sigma_k$.
Indeed, given an $\varepsilon$-ball event that produces the time
series $\tilde \tau_j^i$, in $\Sigma_j$, there is, at least, one
intersection of this ball with the section $\tilde \Gamma_j$.  Since
$\tilde \Gamma_j$ depends on the initial conditions, we can choose an
initial condition right at the $\varepsilon$-ball event.  Next, we
choose $P_{\tilde \Gamma_j}$ such that it is completely covered by the
$\varepsilon$-ball. Since the measure of the $\varepsilon$-ball is
small, $\varepsilon \ll 1$, the time difference between crossings of
the trajectory with $P_{\tilde \Gamma_j}$ and the $\varepsilon$-ball,
namely $\tilde \tau_j^i - \tau_j^{n_i}$, is also small.  Therefore, if
we observe the trajectory of $\Sigma_k$ at times $(\tilde
\tau_j^i)_{i\in \mathbb{N}}$, we get a localized set in $\Sigma_k$
close to the set $\tilde{\mathcal{D}}_k$.  Thus, we conclude our
result.

\begin{figure}[!h]
  \centerline{\hbox{\psfig{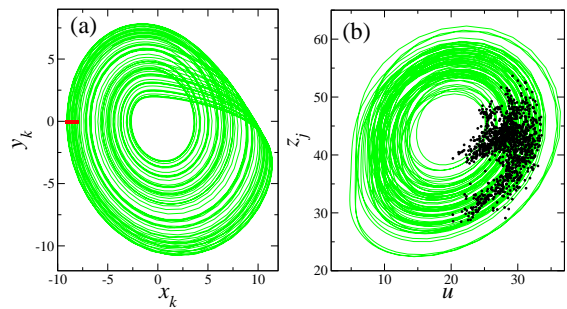}}}
\caption{\label{Nova2} PS implies the existence of localized sets,
  which is generated by the observation of an arbitrary typical event.
  The attractors of the oscillators are depicted in gray.  For
  $\epsilon=13.0$, we show the projections of the R\"ossler Lorenz
  attractors (a,b).  The black bold line shows the segment
  $\mathcal{S}_k$ (a).  $\mathcal{D}_j$ is constructed by observing
  the oscillator $\Sigma_j$ whenever the trajectory of $\Sigma_k$
  crosses the segment $\mathcal{S}_k$. In (b) we show the set
  $\mathcal{D}_j$ in black. Since there is PS, $\mathcal{D}_j$ is
  localized.}
\end{figure}

In order to illustrate these ideas, we consider again the Lorenz
oscillator driven by the R\"ossler oscillator. As we showed before,
for $\epsilon = 13.0$ there is PS. Thus the sets $\mathcal{D}$ must be
localized independently on the event chosen.  We define the event in
the oscillator $\Sigma_k$ to be the crossing of the trajectory
$\mathcal{S}_k = \{x_k,y_k,z_k \in \mathbb{R} | x_j < -8, y_k = 0 ,
\dot{y_k}<0 \}$.  These crossings generate the times $(t_k^i)_{i\in
  \mathbb{N}}$. $\mathcal{S}_k$ is depicted in black bold line in Fig.
\ref{Nova2}(a) together with the attractor of the R\"ossler oscillator
depicted in gray. The observation of the trajectory of $\Sigma_j$ at
the times $(t_k^i)_{i\in \mathbb{N}}$ generates a localized set
$\mathcal{D}_j$[Fig. \ref{Nova2}(b)].

\noindent
\noindent

Keeping these results in mind, we analyze the onset of PS between two
non-coherent neurons of the HR type coupled by chemical synapses.  The
neurons are described by a 4-dimensional HR model \cite{Pinto} which
consists of four coupled differential equations: $\dot{x}_k = ay_k +
bx^2_k - cx^3_k - dz_k + I_k + g_{syn} \sum_j \gamma_{kj}
I_{syn}(x_j),$ $\dot{y}_k = e - y_k + fx^2_k - gw_k,$ $\dot{z}_k =
\mu(-z_k + R(x_k+H)),$$\dot{w}_k = \nu(-kw_k + r(y_k+l)),$ $x_k$
represents the membrane potential, $y_k$ is associated with fast
current dynamics, and $(z_k,w_k)$ are associated with slow currents,
$I_{syn}$ is the synaptic input, and $\gamma_{kj}$ is the connectivity
matrix: $\gamma_{kj} = 1$ if neuron $j$ is connected to neuron $k$,
and $\gamma_{kj} = 0$, otherwise, with $j \not = k$.

We set the parameters of the model in order to obtain a
spiking/bursting behavior \cite{Pinto}. Then, we couple the neurons by
means of chemical synapses.  The current $I_{syn}$ injected in the
postsynaptic cell is given by \cite{Sharp}: $I_{syn}(x_j) = S_j [x_j -
V_{rev}], \tau \frac{dS_j}{dt} = \frac{S_{\infty j }(x_i) - S_j}{S_0 -
  S_{\infty j }(x_i)},$ where $V_{rev}$ is the synaptic potential, and
$\tau$ is the timescale governing receptor binding.  $S_{\infty}$ is
given by: $S_{\infty}(V) = tanh[(V - V_{th})/V_{slope}]$ if $V >
V_{th}$ and $0$ otherwise

\noindent
\begin{figure}[h]
  \centerline{\hbox{\psfig{file=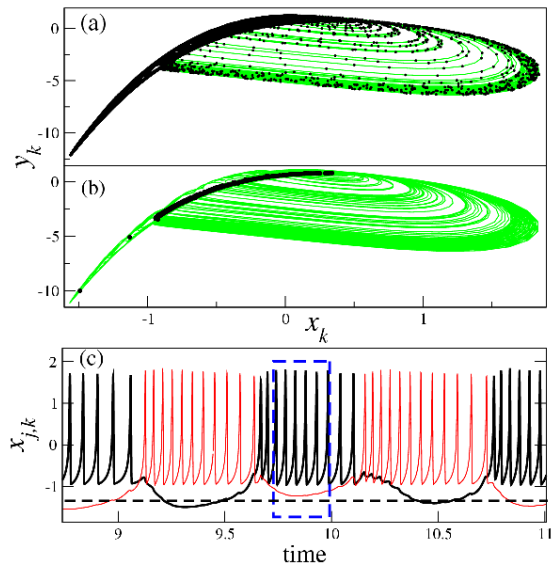,height=8cm}}}
\caption{Onset of PS in two HR neurons coupled via inhibitory synapses. 
  In Figs. (a-b), we plot the attractor projection $(x_k,y_k)$ in
  gray, and the set $\mathcal{D}_k$ in black, constructed by observing
  the subsystem $\Sigma_k$ whenever $\Sigma_j$ crosses the threshold
  $x_j = -1.3$ represented by the dashed line in (c).  In (a) The set
  $\mathcal{D}_k$ spreads over the attractor which shows that there is
  no PS. In (b) the set $\mathcal{D}_k$ is localized which shows the
  presence of PS. In (c) we present the time series of the membrane
  potential of (b). The threshold (dashed line) can mislead the burst
  occurrence (see the box) leading to the wrong statement that there
  is no PS. For (a) the parameters are $I_k = 3.12$ and
  $g_{syn}=0.75$, for (b) and (c) $I_k = 3.12$ and $g_{syn}=0.85$.}
\label{ps-set}
\end{figure}
\noindent
The synapse parameters are chosen in order to have an inhibitory
effect, so, we set: {\small $V_{th}=-0.80$, $V_{slope}=
  1.00$,$V_{rev}=-1.58$, and $S_0 \ge 1$}.

Now, the time series of events $\tau^i_j$ is the time at which the
$i$th crossing of membrane potential $x_j$ reaches a threshold, namely
$x = -1.3$. We fix $I_k=3.1200$ and $I_j=3.1205$, then for $g_{syn} =
0.75$ the set $\mathcal{D}_k$ spreads over the attractor [Fig.
\ref{ps-set}(a)]; there is no PS.  As we increase $g_{syn}$, the
coupled neurons undergo a transition to PS, i.e. the set
$\mathcal{D}_k$ is localized, Fig.  \ref{ps-set}(b). The neurons are
highly non-coherent, due to the existence of two time-scales, and the
inhibitory synapse which causes one neuron that is in a spiking
behavior to inhibit the other neuron, which hyperpolarizes, but still
tries to spike. This competition generates even more non-coherence in
the phase space.  As a consequence, it is rather unclear how one can
calculate the phase for such dynamics. What has been currently done is
to estimate the phase of the chaotic neuron by assuming that in every
crossing in a given direction of the membrane potential with a
threshold, the phase increases $2 \pi$ \cite{reviews}. The main
problem with this approach is that the phase is threshold dependent,
so, it can lead to the false statement that PS does not exist. We
illustrate this problem in Fig \ref{ps-set}(c), for the same parameter
as in Fig \ref{ps-set}(b); there is, indeed, PS.  For a threshold $x =
-1.3$ (dashed line), one burst is missed, what makes the phase
difference to be no longer bounded as the time goes to infinity,
leading to the wrong statement that there is no PS.  Our approach, on
the other hand, is not event dependent.  Indeed, as we showed, a
localized set $\mathcal{D}_k$ exists for this threshold.

Next, we analyze a network of $16$ non-identical HR neurons, connected
with excitatory chemical synapses.  In order to simulate a mismatch in
the intrinsic current, we set $I_i = 3.12 + \eta_i$, where $\eta_i$
are uniformly distributed within the interval $[-0.05,0.05]$.  To
simulate the excitatory synapses, we use the same $I_{syn}$, but
changing the value of $V_{rev}$. Note that if $V_{rev} \ge x_i(t)$ the
neuron presynaptic always injects a positive current in the
postsynaptic neuron. In the following, we set $V_{rev} = 2$.  Our
network is a homogeneous random network, i.e. all neurons receive the
same number $k$ of connections, namely $k=4$.  We constrain $g_{syn}$
to be equal to all neurons. We identify the amount of phase
synchronous neurons by analyzing whether the sets $\mathcal{D}_j$ are
localized, occupying no more than $80\%$ of the attractor of
$\Sigma_j$.

The onset of PS in the whole network takes place at $g^*_{syn} \approx
0.47$, so all neurons become phase synchronized. As the synapse
strength crosses another threshold, $\tilde g_{syn} \approx 0.52$, the
neurons undergo a transition to the rest state, and they no longer
present oscillatory behavior.  Clusters of PS appear even for
$g_{syn}$ far smaller than $g_{syn}^*$.  In fact, right at
$g_{syn}=0.04$, some clusters of neurons exhibit PS among themselves.
These clusters seem to be robust under small perturbations. Clusters
of PS inside the network may offer a suitable environment for
information exchanging.  Each one can be regarded as a channel of
communication, since they possess different frequencies, and therefore
each channel of communication operates in different bandwidths.  This
scenario of cluster formation is neither restricted to this HR model
nor to the synapse model. It can also be found in square-wave and
parabolic bursters.

In conclusion, we have proposed an extension of the stroboscopic map,
as a general way to detect PS in coherent/non-coherent oscillators.
The idea consists in constraining the observation of the trajectory of
an oscillator at the times in which an event occurs in the other
oscillator. We have shown that if PS is present, the maps of the
attractor appear as a localized set in the phase-space, and
vice-versa.  The ideas herein provide a reliable and easy way of
detecting PS, without having to explicitly measure the phase.  This
method can be applied in experiments in real-time and networks.

{\bf Acknowledgment} We thank M. Thiel, M. Romano and C. Zhou for
useful discussions, and Helmholtz Center for Brain and Mind Dynamics
(TP and JK) and Alexander von Humboldt (MSB) for the financial
support.

\end{document}